\documentclass[twocolumn,showpacs,preprintnumbers,superscriptaddress,aps,amsmath,amssymb,pre,10pt]{revtex4-1}

\usepackage{graphicx,easymat,chemarrow}
\usepackage{dcolumn}
\usepackage{bm}
\usepackage{color}
\usepackage{multirow}
\usepackage{listings,url}

\begin{document}

\preprint{{\textit{Physical Review E}} {\textbf{96}} (5), 052201 (2017)}

\title{A direct determination approach for the multifractal\\
        detrending moving average analysis}

\author{Hai-Chuan Xu}
 \affiliation{Research Center for Econophysics, East China University of
   Science and Technology, Shanghai 200237, China}
 \affiliation{Department of Finance, East China University of Science and
   Technology, Shanghai 200237, China}

\author{Gao-Feng Gu}
 \affiliation{Research Center for Econophysics, East China University of
   Science and Technology, Shanghai 200237, China}
 \affiliation{Department of Finance, East China University of Science and
   Technology, Shanghai 200237, China}


\author{Wei-Xing Zhou}
 \email{wxzhou@ecust.edu.cn}
 \affiliation{Research Center for Econophysics, East China University of
   Science and Technology, Shanghai 200237, China}
 \affiliation{Department of Finance, East China University of Science and
   Technology, Shanghai 200237, China}
 \affiliation{School of Science, East China University of Science and
   Technology, Shanghai 200237, China}


\date{\today}

\begin{abstract}
  In the canonical framework, we propose an alternative approach for the multifractal analysis based on the detrending moving average method (MF-DMA). We define a canonical measure such that the multifractal mass exponent $\tau(q)$ is related to the partition function and the multifractal spectrum $f(\alpha)$ can be directly determined. The performances of the direct determination approach and the traditional approach of the MF-DMA are compared based on three synthetic multifractal and monofractal measures generated from the one-dimensional $p$-model, the two-dimensional $p$-model and the fractional Brownian motions. We find that both approaches have comparable performances to unveil the fractal and multifractal nature. In other words, without loss of accuracy, the multifractal spectrum $f(\alpha)$ can be directly determined using the new approach with less computation cost. We also apply the new MF-DMA approach to the volatility time series of stock prices and confirm the presence of multifractality.
\end{abstract}

\pacs{89.75.Da, 05.45.Tp, 05.45.Df, 05.40.-a}

\maketitle

\section{Introduction}

The long-range behavior of many chaotic, nonlinear dynamical systems can be described by fractal or multifractal measures  \cite{Mandelbrot-1983,Mandelbrot-1997,Sornette-2004}. A large number of methods have been proposed to characterize the properties of fractals and multifractals. One of the most classic methods is the Hurst analysis or rescaled range analysis (R/S)  \cite{Hurst-1951-TASCE,Mandelbrot-Wallis-1969b-WRR}. The wavelet transform module maxima (WTMM) approach is also a powerful tool \cite{Holschneider-1988-JSP,Muzy-Bacry-Arneodo-1991-PRL,Muzy-Bacry-Arneodo-1993-PRE,Bacry-Muzy-Arneodo-1993-JSP,Muzy-Bacry-Arneodo-1994-IJBC}, even for high-dimensional multifractal measures, e.g. image technology and turbulence \cite{Arneodo-Decoster-Roux-2000-EPJB,Decoster-Roux-Arneodo-2000-EPJB,Roux-Arneodo-Decoster-2000-EPJB,Kestener-Arneodo-2003-PRL,Kestener-Arneodo-2004-PRL}.
Another popular family include the detrended fluctuation analysis (DFA) \cite{Peng-Buldyrev-Havlin-Simons-Stanley-Goldberger-1994-PRE,Kantelhardt-Zschiegner-KoscielnyBunde-Havlin-Bunde-Stanley-2002-PA,Kristoufek-2015-PRE} and the detrending moving average analysis (DMA) \cite{Alessio-Carbone-Castelli-Frappietro-2002-EPJB,Arianos-Carbone-2007-PA,Carbone-2009-IEEE,Carbone-Kiyono-2016-PRE}. Extensive numerical simulations display that the performance of the DMA approach is comparable to the DFA approach with slightly different priorities under different situations \cite{Xu-Ivanov-Hu-Chen-Carbone-Stanley-2005-PRE,Grech-Mazur-2005-APPB,Oswiecimka-Kwapien-Drozdz-2006-PRE,Bashan-Bartsch-Kantelhardt-Havlin-2008-PA,Gu-Zhou-2010-PRE,Shao-Gu-Jiang-Zhou-Sornette-2012-SR,Shao-Gu-Jiang-Zhou-2015-Fractals,Kiyono-Tsujimoto-2016-PRE}. In real applications, one should keep it in mind that the determination of scaling ranges plays a crucial role in computing the scaling exponents \cite{Grech-Mazur-2013-PA,Grech-Mazur-2013-PRE,Grech-Mazur-2015-APPA}. These methods have been extended to many directions, such as objects in high dimensions \cite{Gu-Zhou-2006-PRE,Carbone-2007-PRE,AlvarezRamirez-Echeverria-Rodriguez-2008-PA,Turk-Carbone-Chiaia-2010-PRE,Tsujimoto-Miki-Shimatani-Kiyono-2016-PRE}, detrended cross-correlation analysis and its variants for two time analysis \cite{Meneveau-Sreenivasan-Kailasnath-Fan-1990-PRA,Jun-Oh-Kim-2006-PRE,Podobnik-Stanley-2008-PRL,Zhou-2008-PRE,Podobnik-Horvatic-Petersen-Stanley-2009-PNAS,Horvatic-Stanley-Podobnik-2011-EPL,Jiang-Zhou-2011-PRE,Kristoufek-2011-EPL, Wang-Shang-Ge-2012-Fractals,Oswiecimka-Drozdz-Forczek-Jadach-Kwapien-2014-PRE,Kwapien-Oswiecimka-Drozdz-2015-PRE,Xie-Jiang-Gu-Xiong-Zhou-2015-NJP,Jiang-Yang-Wang-Zhou-2017-FoP,Jiang-Gao-Zhou-Stanley-2017-Fractals}, detrended partial cross-correlation analysis for multivariate time series \cite{Liu-2014,Yuan-Fu-Zhang-Piao-Xoplaki-Luterbacher-2015-SR,Qian-Liu-Jiang-Podobnik-Zhou-Stanley-2015-PRE}, and so on.

In this paper, inspired by the idea of direct determination of the $f(\alpha)$ singularity spectrum through canonical measures in the partition function approach \cite{Chhabra-Jensen-1989-PRL,Chhabra-Meneveau-Jensen-Sreenivasan-1989-PRA,Meneveau-Sreenivasan-Kailasnath-Fan-1990-PRA}, we develop the MF-DMA approach by defining a canonical measure such that the singularity strength function $\alpha(q)$ and the multifractal spectrum $f(\alpha)$ can also be directly determined. The original MF-DMA approach \cite{Gu-Zhou-2010-PRE,Schumann-Kantelhardt-2011-PA} requires that the Hurst exponent $h(q)$ should first be calculated, then multifractal scaling exponent $\tau(q)$, finally $\alpha(q)$ and $f(\alpha)$ via the Legendre transform \cite{Halsey-Jensen-Kadanoff-Procaccia-Shraiman-1986-PRA}. The modified approach is designed to analyze multifractal time series and multifractal surfaces. The performances of this new MF-DMA approach are investigated using synthetic fractal and multifractal measures with known scaling properties.

The paper is organized as follows. In Sec. \ref{S1:MF-DMA-ONE}, we describe the direct determination approach and the traditional approach for MF-DMA. The one-dimensional and two-dimensional cases are presented separately. In Sec. \ref{S1:Numerical}, we compare the performances of these two approaches through numerical simulations. We consider three numerical experiments, i.e., one-dimensional $p$-model, two-dimensional $p$-model and fractional Brownian motion. In Sec. \ref{S1:Application}, we apply the MF-DMA approaches to analyze time series of intraday stock returns. We discuss and conclude in Sec. \ref{S1:Conclude}.

\section{Multifractal detrending moving average analysis}
\label{S1:MF-DMA-ONE}

In this section, for both one-dimensional case and two-dimensional case, we first present the new direct determination approach, and then describe the traditional approach of MF-DMA analysis \cite{Gu-Zhou-2010-PRE,Schumann-Kantelhardt-2011-PA}.

\subsection{One-dimensional case: MF-DMA$(\theta,q)$}


Consider a time series $x(t)$, $t=1,2,\cdots, N$. We construct the sequence of cumulative sums
\begin{equation}
y(t)=\sum_{i=1}^{t}{x(i)}, ~~t=1, 2, \cdots, N.
 \label{Eq:MF-DMA-OG_y}
\end{equation}
The moving average function $\widetilde{y}(t)$ in a moving window can be calculated as follows \cite{Arianos-Carbone-2007-PA},
\begin{equation}
\widetilde{y}(t)=\frac{1}{s}\sum_{k=-\lfloor(s-1)\theta\rfloor}^{\lceil(s-1)(1-\theta)\rceil}y(t-k),
\label{Eq:MF-DMA-OG_y1}
\end{equation}
where $s$ is the window size, $\lfloor{x}\rfloor$ is the largest integer not greater than $x$, $\lceil{x}\rceil$ is the smallest integer not smaller than $x$, and $\theta$ is the position parameter with the value varying in the range $[0,1]$. The cases $\theta=0$, $\theta=0.5$ and $\theta=1$ refer respectively to the backward, centred and forward moving average analysis \cite{Xu-Ivanov-Hu-Chen-Carbone-Stanley-2005-PRE}. The trend $\widetilde{y}(t)$ can also be estimated by higher order polynomials \cite{Arianos-Carbone-Turk-2011-PRE}, however, the implementation of higher order DMA significantly increases the computational cost, which would prevent practical use of this method.

We detrend the signal series by removing the moving average function $\widetilde{y}(i)$ from $y(i)$, and obtain the residual sequence $\epsilon(i)$ through
\begin{equation}
\epsilon(i)=y(i)-\widetilde{y}(i),
\label{Eq:MF-DMA-OG:epsilon}
\end{equation}
where $s-\lfloor(s-1)\theta\rfloor\leqslant{i}\leqslant{N-\lfloor(s-1)\theta\rfloor}$. The residual series $\epsilon(i)$ is divided into $N_s$ disjoint sub-series with the same size $s$, where $N_s=\lfloor{N}/s-1\rfloor$. Each sub-series can be denoted by $\epsilon_v$ such that $\epsilon_v(i)=\epsilon(l+i)$ for $1\leqslant{i}\leqslant{s}$, where $l=(v-1)s$. We calculate the root-mean-square function $F_v(s)$ as follows
\begin{equation}
F_v(s)=\left[\frac{1}{s}\sum_{i=1}^{s}\epsilon_v^2(i)\right]^{1/2}.
\label{Eq:MF-DMA-OG:Fv}
\end{equation}

The function $F_v(s)$ reflects the amount of the residual sequence within each segment $v$ of size $s$, which is known as the box probability in the standard textbook box counting formalism. From the canonical perspective, one can obtain the $f(\alpha)$ function directly \cite{Chhabra-Jensen-1989-PRL,Chhabra-Meneveau-Jensen-Sreenivasan-1989-PRA,Meneveau-Sreenivasan-Kailasnath-Fan-1990-PRA}. Here we define a canonical measure $\mu(q,s,v)$ using the fluctuation function $F_v(s)$:
\begin{equation}
  \mu(q,s,v) = \frac{F_v^q(s)}{\sum_{v=1}^{N_s} F_v^q(s)},
  \label{Eq:MF-DMA-CN:mu}
\end{equation}
where $q$ is the index variable. Let the partition function $\chi(q,s)= \sum_{v=1}^{N_s} F_v^q(s)$, from which can obtain the multifractal mass exponent $\tau(q)$, that is,
\begin{equation}
  \chi(q,s) \sim s^{\tau(q)}.
  \label{Eq:MF-DMA-PF:chi:tau}
\end{equation}

Then, the singularity strength $\alpha(q)$ and the multifractal spectrum $f(\alpha)$ are related to $\tau(q)$ via a Legendre transform. Substituting partition function $\chi(q,s)$ and canonical measure $\mu(q,s,v)$, $\alpha(q)$ and $f(\alpha)$ are deduced as
\begin{subequations}
\begin{equation}
\begin{split}
  \alpha(q) & = \frac{{\rm{d}}\tau(q)}{{\rm{d}}q} = \lim_{s\to0} \frac{{\rm{d}}}{{\rm{d}}q} \frac{\ln \chi(q,s)}{\ln{s}} \\
            & = \lim_{s\to0} \frac{\sum_{v=1}^{N_s} F_v^q(s) \ln F_v(s)}{\sum_{v=1}^{N_s}F_v^q(s)\ln{s}} \\
            & = \lim_{s\to0} \frac{\sum_{v=1}^{N_s} \mu(q,s,v) \ln F_v(s)}{\ln{s}},
\end{split}
  \label{Eq:MF-DMA-CN:alpha}
\end{equation}
and
\begin{equation}
\begin{split}
  & f(\alpha(q)) = q\alpha(q)-\tau(q) \\
  & = \lim_{s\to0}\frac{q\sum_{v=1}^{N_s}[F_v^q(s)\ln F_v(s)]-\sum_{v=1}^{N_s}F_v^q(s)\ln[\sum_{v=1}^{N_s}F_v^q(s)]}{\sum_{v=1}^{N_s}F_v^q(s)\ln{s}} \\
  & = \lim_{s\to0}\frac{\sum_{v=1}^{N_s}F_v^q(s)[\ln{F_v^q(s)}-\ln[\sum_{v=1}^{N_s}F_v^q(s)]]}{\sum_{v=1}^{N_s}F_v^q(s)\ln{s}}\\
  & = \lim_{s\to0}\frac{\sum_{v=1}^{N_s} \mu(q,s,v) \ln\left[\mu(q,s,v)\right]}{\ln{s}}.
\end{split}
  \label{Eq:MF-DMA-CN:fq}
\end{equation}
\label{Eq:MF-DMA-CN:alpha:fq}
\end{subequations}
In practice, $\alpha(q)$ and $f(\alpha)$ can be computed by linear regressions in semi-log coordinates. The multifractal spectrum $f(\alpha)$ is thus directly determined by the measure $\mu(q,s,v)$. That is, Eq.~(\ref{Eq:MF-DMA-CN:mu}) and Eq.~(\ref{Eq:MF-DMA-CN:alpha:fq}) are the ``canonical'' counterparts of the original MF-DMA method \cite{Gu-Zhou-2010-PRE}.


%

%

In the traditional MF-DMA analysis, the $q$th order overall fluctuation function $F(q,s)$ is calculated as follows,
\begin{equation}
  F(q,s) = \left\{\frac{1}{N_s}\sum_{v=1}^{N_s} {F_v^q(s)}\right\}^{\frac{1}{q}},
  \label{Eq:MF-DMA-OG:Fqs}
\end{equation}
where $q$ can take any real value except for $q=0$. When $q=0$, we have
\begin{equation}
  \ln[F(0,s)] = \frac{1}{N_s}\sum_{v=1}^{N_s}{\ln[F_v(s)]},
  \label{Eq:MF-DMA-OG:Fq0}
\end{equation}
according to L'H\^{o}spital's rule. Varying the values of $s$, we can determine the power-law relation between the function $F(q,s)$ and the size scale $s$:
\begin{equation}
  F(q,s)\sim{s}^{h(q)}.
  \label{Eq:MF-DMA-OG:Fqs:h}
\end{equation}
The multifractal scaling exponent $\tau(q)$ can be be determined as follows
\begin{equation}
\tau(q)=qh(q)-D_f,
\label{Eq:MF-DMA-OG:tau:hq}
\end{equation}
where $D_f$ is the fractal dimension of the geometric support of the multifractal measure. If the scaling exponent function $\tau(q)$ is a nonlinear function of $q$, the signal has multifractal nature. The order-$q$ generalized dimension $D_q$ can be obtained by
\begin{equation}
  D_q = \frac{\tau(q)}{q-1}.
\end{equation}
Based on the Legendre transform, we can obtain the singularity strength function $\alpha(q)$ and the multifractal spectrum $f(\alpha)$ \cite{Halsey-Jensen-Kadanoff-Procaccia-Shraiman-1986-PRA}
\begin{subequations}
\begin{equation}
  \alpha(q) = \frac{{\rm{d}}\tau(q)}{{\rm{d}}q} = h(q) + q \frac{{\rm{d}}h(q)}{{\rm{d}}q},
  \label{Eq:MF-DMA-OG:alpha}
\end{equation}
and
\begin{equation}
  f(\alpha(q)) = q{\alpha}-{\tau}(q) = q[\alpha - h(q)] + D_f.
  \label{Eq:MF-DMA-OG:fq}
\end{equation}
\label{Eq:MF-DMA-OG:alpha:fq}
\end{subequations}

\begin{figure*}[htb]
  \centering
  \includegraphics[width=0.9\linewidth]{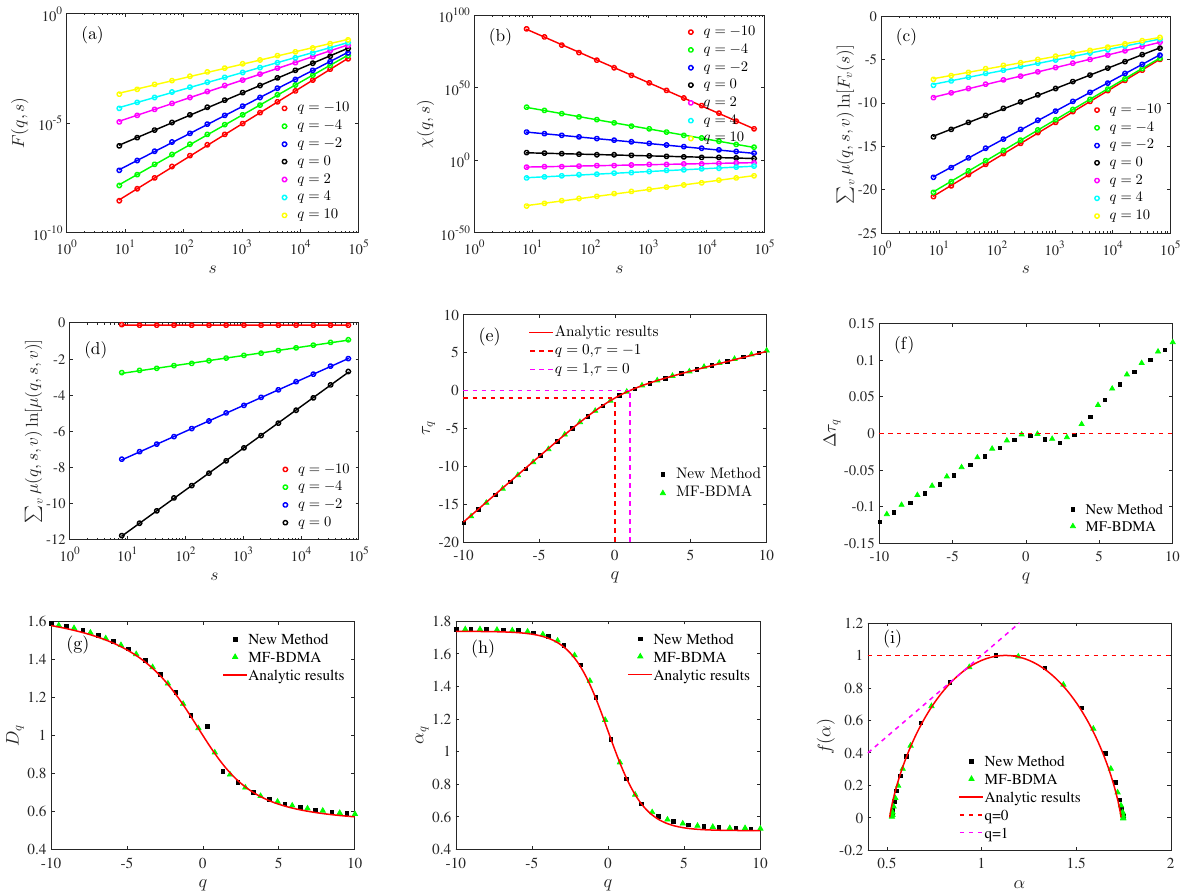}
  \caption{Multifractal analysis of a binomial measure with $p=0.3$, based on the traditional MF-BDMA$(\theta,q)$ and the direct determination approach with $\theta=0$. (a) Power-law dependence of $F(q,s)$ on box size $s$ for different $q$. (b) Power-law dependence of $\chi(q,s)$ on box size $s$ for different $q$. (c) Linear dependence of $\sum_v \mu(q,s,v) \ln{F_v(s)}$ against $\ln{s}$. (d) Linear dependence of $\sum_v\mu(q,s,v)\ln[\mu(q,s,v)]$ against $\ln{s}$. (e) The mass exponent function $\tau(q)$. (f) Differences $\Delta \tau(q)$ between the estimated mass exponents and their theoretical values. (g) The generalized dimensions $D_q$. (h) The singularity strength function $\alpha(q)$. (i) The multifractal singularity spectrum $f(\alpha)$.}
  \label{Fig:MF-DMA:p-model:theta=0}
\end{figure*}

\subsection{Two-dimensional case: MF-DMA$(\theta_1,\theta_2,q)$}



The two-dimensional MF-DMA analysis is used to investigate possible multifractal properties of surfaces $X(i_1,i_2)$ with $i_1=1,2,\cdots,N_1$ and $i_2=1,2,\cdots,N_2$. Some surface analyses (e.g. fractal cracks) measure two independent (or dependent) scaling exponents along the front direction and along the propagation direction, in the sense that the local height (out-of-plane) scales as the in-plane displacement in two separate directions\cite{Ponson-Bonamy-Bouchaud-2006-PRL}. Differently, what we concerned here is the scaling behavior on the partitioned squares, not on a directed displacement. The local detrended fluctuation $F_{v_1,v_2}(s_1,s_2)$ can be calculated as follows,
\begin{equation}
  F_{v_1,v_2}(s_1,s_2)=\left[\frac{1}{s_1s_2}\sum_{i_1=1}^{s_1}\sum_{i_2=1}^{s_2}\epsilon_{v_1,v_2}^2(i_1,i_2)\right]^{1/2}.
  \label{Eq:MF-DMA-OG:Fv:2D}
\end{equation}
The residual matrix $\epsilon(i_1,i_2)$ is partitioned into $N_{s_1}\times{N_{s_2}}$ disjoint rectangle segments of the same size $s_1\times{s_2}$, where $N_{s_1}=\lfloor{(N_1-s_1(1+\theta_1))/s_1}\rfloor$ and $N_{s_2}=\lfloor{(N_2-s_2(1+\theta_2))/s_2}\rfloor$. Each segment is denoted by $\epsilon_{v_1,v_2}$ such that $\epsilon_{v_1,v_2}(i_1,i_2)=\epsilon(l_1+i_1,l_2+i_2)$ for $1\leqslant{i_1}\leqslant{s_1}$ and $1\leqslant{i_2}\leqslant{s_2}$, where $l_1=(v_1-1)s_1$ and $l_2=(v_2-1)s_2$.

Generally, we set $s=s_1=s_2$. Similar to the one-dimensional case, we define the canonical measures as
\begin{equation}
  \mu(q,s,v_1,v_2) = \frac{F_{v_1,v_2}^{q}(s)}{\sum_{v_1}\sum_{v_2}F_{v_1,v_2}^{q}(s)},
  \label{Eq:MF-DMA-CN:mu:2D}
\end{equation}
Then, from the partition function $\chi(q,s)$ we can obtain the multifractal mass exponent $\tau(q)$, that is,
\begin{equation}
  \chi(q,s)= \sum_{v_1}\sum_{v_2} F_{v_1,v_2}^{q}(s) \sim s^{\tau(q)}.
  \label{Eq:MF-DMA-PF:chi:def:2D}
\end{equation}
Similar to the one-dimensional case, the singularity strength $\alpha(q)$ and the singularity spectrum $f(\alpha)$ are deduced as
\begin{subequations}
\begin{equation}
\begin{split}
  \alpha(q) & = \frac{{\rm{d}}\tau(q)}{{\rm{d}}q} = \lim_{s\to0} \frac{{\rm{d}}}{{\rm{d}}q} \frac{\ln \chi(q,s)}{\ln{s}} \\
            & = \lim_{s\to0} \frac{\sum_{v_1}\sum_{v_2} F_{v_1,v_2}^{q}(s) \ln{F_{v_1,v_2}(s)}}{\sum_{v_1}\sum_{v_2} F_{v_1,v_2}^{q}(s)\ln{s}} \\
            & = \lim_{s\to0} \frac{\sum_{v_1}\sum_{v_2} \mu(q,s,v_1,v_2) \ln{F_{v_1,v_2}(s)}}{\ln{s}},
\end{split}
  \label{Eq:MF-DMA-CN:alpha:2D}
\end{equation}
and
\begin{equation}
\hspace{-0.4in}
\begin{split}
  & f(\alpha(q)) = q\alpha(q)-\tau(q) \\
  & = \lim_{s\to0}\left\{\frac{q\sum_{v_1}\sum_{v_2}[F_{v_1,v_2}^{q}(s)\ln{F_{v_1,v_2}(s)}]}
  {\sum_{v_1}\sum_{v_2}F_{v_1,v_2}^{q}(s)\ln{s}}-\frac{\ln[\sum_{v_1}\sum_{v_2}F_{v_1,v_2}^{q}(s)]}{\ln{s}}\right\} \\
  & = \lim_{s\to0}\frac{\sum_{v_1}\sum_{v_2}F_{v_1,v_2}^{q}(s)[\ln{F_{v_1,v_2}^{q}(s)}
  -\ln[\sum_{v_1}\sum_{v_2}F_{v_1,v_2}^{q}(s)]]}{\sum_{v_1}\sum_{v_2}F_{v_1,v_2}^{q}(s)\ln{s}}\\
  & = \lim_{s\to0}\frac{\sum_{v_1}\sum_{v_2} \mu(q,s,v_1,v_2) \ln\left[\mu(q,s,v_1,v_2)\right]}{\ln{s}}.
\end{split}
\hspace{+0.2in} \label{Eq:MF-DMA-CN:fq:2D}
\end{equation}
\label{Eq:MF-DMA-CN:alpha:fq:2D}
\end{subequations}
In practice, $\alpha(q)$ and $f(\alpha)$ can be computed by linear regressions in semi-log scales.


\begin{figure*}[htb]
  \centering
  \includegraphics[width=0.9\linewidth]{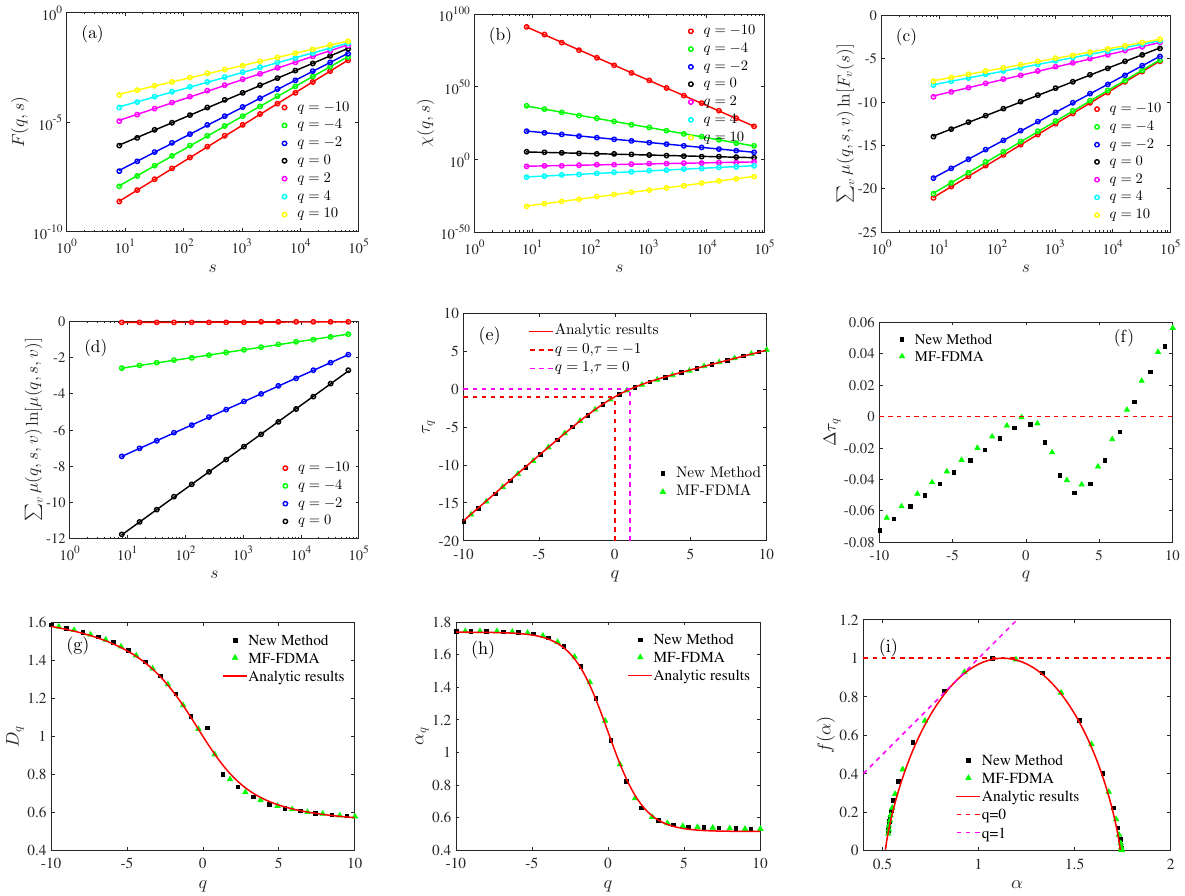}
  \caption{Multifractal analysis of a binomial measure with $p=0.3$, based on the traditional MF-FDMA$(\theta,q)$ and the direct determination approach with $\theta=1$. (a) Power-law dependence of $F(q,s)$ on box size $s$ for different $q$. (b) Power-law dependence of $\chi(q,s)$ on box size $s$ for different $q$. (c) Linear dependence of $\sum_v \mu(q,s,v) \ln{F_v(s)}$ against $\ln{s}$. (d) Linear dependence of $\sum_v\mu(q,s,v)\ln[\mu(q,s,v)]$ against $\ln{s}$. (e) The mass exponent function $\tau(q)$. (f) Differences $\Delta \tau(q)$ between the estimated mass exponents and their theoretical values. (g) The generalized dimensions $D_q$. (h) The singularity strength function $\alpha(q)$. (i) The multifractal singularity spectrum $f(\alpha)$.}
  \label{Fig:MF-DMA:p-model:theta=1}
\end{figure*}

In the original two-dimensional MF-DMA method\cite{Gu-Zhou-2010-PRE}, the $q$th order overall fluctuation function $F(q,s)$ is calculated as follows,
\begin{equation}
  F(q,s) = \left\{\frac{1}{N_{s_1}N_{s_2}}\sum_{v_1=1}^{N_{s_1}}\sum_{v_2=1}^{N_{s_2}}{F_{v_1,v_2}^q(s_1,s_2)}\right\}^{\frac{1}{q}},
  \label{Eq:MF-DMA-OG:Fq:2D}
\end{equation}
where $q$ can take any real values except for $q=0$. When $q=0$, we have
\begin{equation}
  \ln[F(0,s)] = \frac{1}{N_{s_1}N_{s_2}}\sum_{v_1=1}^{N_{s_1}}\sum_{v_2=1}^{N_{s_2}}{\ln[F_{v_1,v_2}(s_1,s_2)]},
  \label{Eq:MF-DMA-OG:F0:2D}
\end{equation}
according to L'H\^{o}spital's rule.
Varying the segment sizes $s_1$ and $s_2$, we are able to determine the power-law relation between the fluctuation function ${F(q,s)}$ and the scale $s$,
\begin{equation}
  F(q,s)\sim{s}^{h(q)},
  \label{Eq:MF-DMA-OG:h:2D}
\end{equation}
Applying Eqs.~(\ref{Eq:MF-DMA-OG:tau:hq}) and (\ref{Eq:MF-DMA-OG:alpha:fq}), we can obtain the multifractal scaling exponent $\tau(q)$, the singularity strength function $\alpha(q)$ and the multifractal spectrum $f(\alpha)$, respectively. For two-dimensional multifractal measures, we have $D_f=2$ in Eq.~(\ref{Eq:MF-DMA-OG:tau:hq}).

\section{Numerical experiments}
\label{S1:Numerical}

\subsection{One-dimensional $p$-model}

\begin{figure*}[htb]
  \centering
  \includegraphics[width=0.9\linewidth]{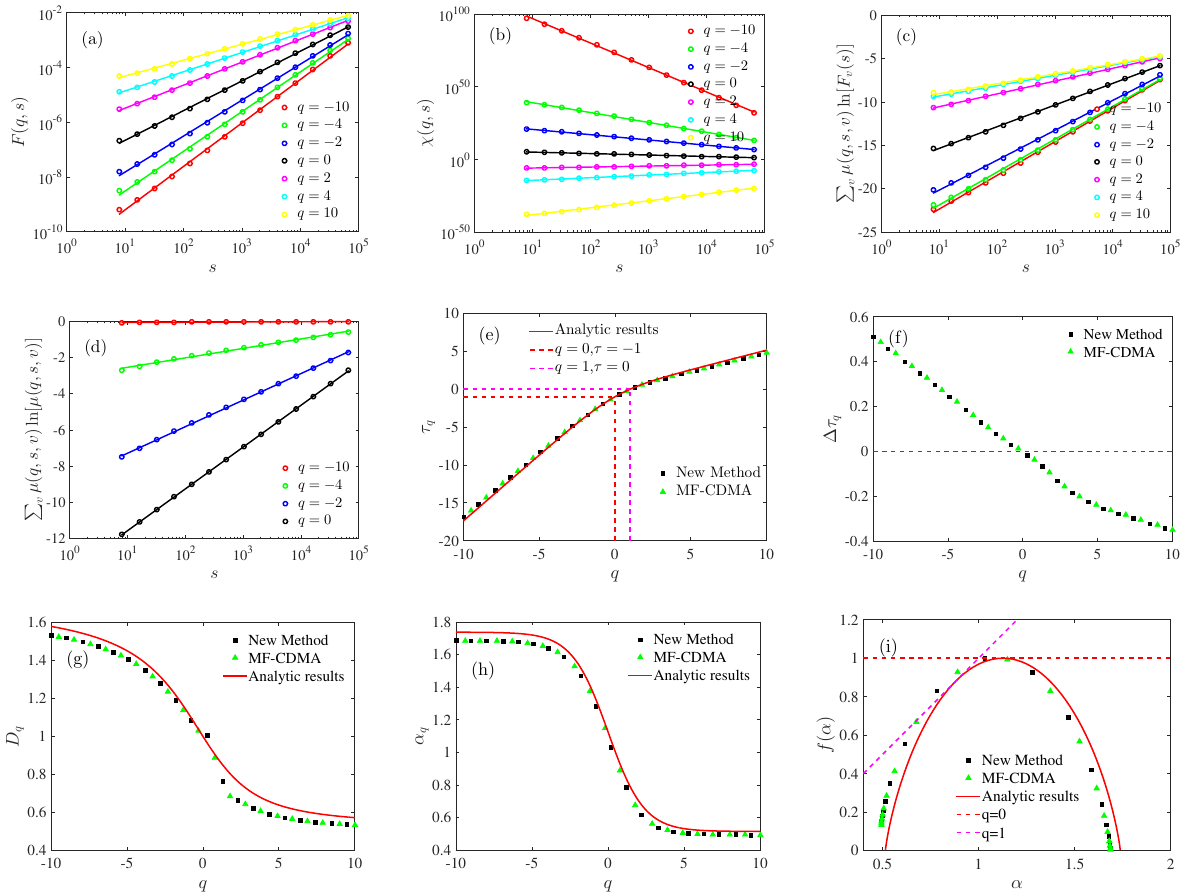}
  \caption{Multifractal analysis of a binomial measure with $p=0.3$, based on the traditional MF-CDMA$(\theta,q)$ and the direct determination approach with $\theta=0.5$. (a) Power-law dependence of $F(q,s)$ on box size $s$ for different $q$. (b) Power-law dependence of $\chi(q,s)$ on box size $s$ for different $q$. (c) Linear dependence of $\sum_v \mu(q,s,v) \ln{F_v(s)}$ against $\ln{s}$. (d) Linear dependence of $\sum_v\mu(q,s,v)\ln[\mu(q,s,v)]$ against $\ln{s}$. (e) The mass exponent function $\tau(q)$. (f) Differences $\Delta \tau(q)$ between the estimated mass exponents and their theoretical values. (g) The generalized dimensions $D_q$. (h) The singularity strength function $\alpha(q)$. (i) The multifractal singularity spectrum $f(\alpha)$.}
  \label{Fig:MF-DMA:p-model:theta=05}
\end{figure*}

To investigate the performance of different MF-DMA approaches, we apply the $p$-model \cite{Meneveau-Sreenivasan-1987-PRL} to synthesize multifractal time series. The $p$-model can produce standard multifractal series and thus the analytical formula of the scaling exponents $\tau(q)$ and the singularity strength function $\alpha(q)$ can be obtained exactly. Therefore, the $p$-model is used to test the performance of the multifractal estimators \cite{Kantelhardt-Zschiegner-KoscielnyBunde-Havlin-Bunde-Stanley-2002-PA,Gu-Zhou-2010-PRE}. Start from a measure $m$ uniformly distributed on an interval $[0,1]$. In the first step, the measure is redistributed on the interval, $m_{1,1}=mp_1$ to the first half interval and $m_{1,2}=mp_2=m(1-p_1)$ to the second half interval. In the $(k+1)$-th step, the measure $m_{k,i}$ on each of the $2^k$ line segments is redistributed into two parts, where $m_{k+1,2i-1}=m_{k,i}p_1$ and  $m_{k+1,2i}=m_{k,i}p_2$. We repeat the procedure for $20$ times and finally generate the multifractal time series with a length of $2^{20}=1048576$. We present the results when the parameters are $p_1=0.3$ and $p_2=0.7$ and compare the performances of the backward moving average ($\theta=0$), the centered moving average ($\theta=0.5$) and the forward moving average ($\theta=1$). The results for other parameters are qualitatively the same.


We elaborate on the case of backward moving average in Fig.~\ref{Fig:MF-DMA:p-model:theta=0}. Fig.~\ref{Fig:MF-DMA:p-model:theta=0}(a) illustrates the power-law dependence of the fluctuation function $F(q,s)$ on the scale $s$ for different $q$. The exponents $h(q)$ for the traditional MF-DMA method are obtained by the least squares fitting in log-log scales. Fig.~\ref{Fig:MF-DMA:p-model:theta=0}(b) illustrates the power-law dependence of the partition function $\chi(q,s)$ on the scale $s$ for different $q$. The slopes obtained by linear regressions of $\ln \chi(q,s)$ against $\ln s$ are the estimates of $\tau(q)$, which are shown in Fig.~\ref{Fig:MF-DMA:p-model:theta=0}(e). We obtain $D_q$ using $D_q = \frac{\tau(q)}{q-1}$ and $\alpha(q)$ and $f(\alpha)$ using the Legendre transform, which are presented in Fig.~\ref{Fig:MF-DMA:p-model:theta=0}(g-i). On the other hands, the multifractal nature in the one-dimensional $p$-model can also be estimated by the direct determination approach.
Fig.~\ref{Fig:MF-DMA:p-model:theta=0}(c) plots the dependence of $\sum_{v=1}^{N_s} \mu(q,s,v) \ln F_v(s)$ against $s$ and Fig.~\ref{Fig:MF-DMA:p-model:theta=0}(d) plots the dependence of $\sum_{v=1}^{N_s} \mu(q,s,v) \ln\left[\mu(q,s,v)\right]$ against $s$ in linear-log coordinates. The slopes of the linear fits in Fig.~\ref{Fig:MF-DMA:p-model:theta=0}(c) and Fig.~\ref{Fig:MF-DMA:p-model:theta=0}(d) are the estimates of $\alpha(q)$ and $f(\alpha)$ directly, which are shown in Fig.~\ref{Fig:MF-DMA:p-model:theta=0}(h-i). In Fig.~\ref{Fig:MF-DMA:p-model:theta=0}(e) and Fig.~\ref{Fig:MF-DMA:p-model:theta=0}(g-i), we also show the analytical solution as a continuous curve for comparison. The analytical formula of $\tau(q)$ for time series generated by the $p$-model can be expressed by \cite{Halsey-Jensen-Kadanoff-Procaccia-Shraiman-1986-PRA},
\begin{equation}
  \tau_{\rm{analy}}(q)=-\frac{\ln(p_1^q+p_2^q)}{\ln2}.
  \label{Eq:tau:analy}
\end{equation}
The analytical singularity strength function $\alpha(q)$ can be calculated as follows
\begin{equation}
  \alpha_{\rm{analy}}(q)=-\frac{p_1^q\ln{p_1}+p_2^q\ln{p_2}}{(p_1^q+p_2^q)\ln2}.
  \label{Eq:alpha:analy}
\end{equation}
The analytical expressions of $D_q$ and the multifractal spectrum $f(\alpha)$ can be obtained via $D_q = \frac{\tau(q)}{q-1}$ and $f(\alpha) = q\alpha - \tau(q)$ respectively. In Fig.~\ref{Fig:MF-DMA:p-model:theta=0}(e), we mark $\tau(0)=-1$ and $\tau(1)=0$, while in Fig.~\ref{Fig:MF-DMA:p-model:theta=0}(i), we show that $f^\prime(\alpha\mid q=0) = 0$ and $f^\prime(\alpha\mid q=1) = 1$.
It is evident that both the direct determination approach and the traditional approach can unveil the multifractal nature of the binomial measure with very high accuracy.

To make more precise description, we display the differences $\Delta\tau(q)$ between the estimated mass exponents and their theoretical values in Fig.~\ref{Fig:MF-DMA:p-model:theta=0}(f). we find that the estimation deviations became bigger when the $q$ values are far from 0. Very negative $q$'s cause the mass exponents $\tau(q)$ underestimated and very positive $q$'s cause $\tau(q)$ overestimated. Correspondingly, when $|q|$'s are setting large, the estimated $D_q$ in Fig.~\ref{Fig:MF-DMA:p-model:theta=0}(g) and $\alpha(q)$ in Fig.~\ref{Fig:MF-DMA:p-model:theta=0}(h) are slightly higher than their corresponding analytic values, and the estimated $f(\alpha)$ curve in Fig.~\ref{Fig:MF-DMA:p-model:theta=0}(i) is slightly right-biased to their analytic values. Another property observed in Fig.~\ref{Fig:MF-DMA:p-model:theta=0}(f) is that the $\Delta\tau(q)$ function estimated by the direct determination approach locates slightly lower than the estimation of traditional approach. This indicates that, for very positive $q$'s, the direct determination approach performs better than the traditional MF-DMA analysis. However, for very negative $q$'s, the traditional approach performs better.

The case of forward moving average displayed in Fig.~\ref{Fig:MF-DMA:p-model:theta=1} is very similar with the case of backward moving average. However, the case of centered moving average displayed in Fig.~\ref{Fig:MF-DMA:p-model:theta=05} has something different. In Fig.~\ref{Fig:MF-DMA:p-model:theta=05}(f), very negative $q$'s cause the mass exponents $\tau(q)$ overestimated and very positive $q$'s cause $\tau(q)$ underestimated. Correspondingly, when $|q|$ are setting large, the estimated $D_q$ in Fig.~\ref{Fig:MF-DMA:p-model:theta=05}(g) and $\alpha(q)$ in Fig.~\ref{Fig:MF-DMA:p-model:theta=05}(h) are slightly lower than their respective analytic values, and the estimated $f(\alpha)$ in Fig.~\ref{Fig:MF-DMA:p-model:theta=05}(i) are slightly left-biased to their analytic values. We notice that $\Delta\tau(q)$ estimated by the direct determination approach locate almost the same as the estimation of traditional approach. Therefore, the performances of MF-CDMA analysis done by these two approaches are comparable. Here we stress that, no matter which approach we used, both the backward and the forward MF-DMA methods outperform the centered MF-DMA method, when considering the difference $\Delta\tau(q)$.

\subsection{Two-dimensional $p$-model}

\begin{figure}[htb]
  \centering
  \includegraphics[width=0.9\linewidth]{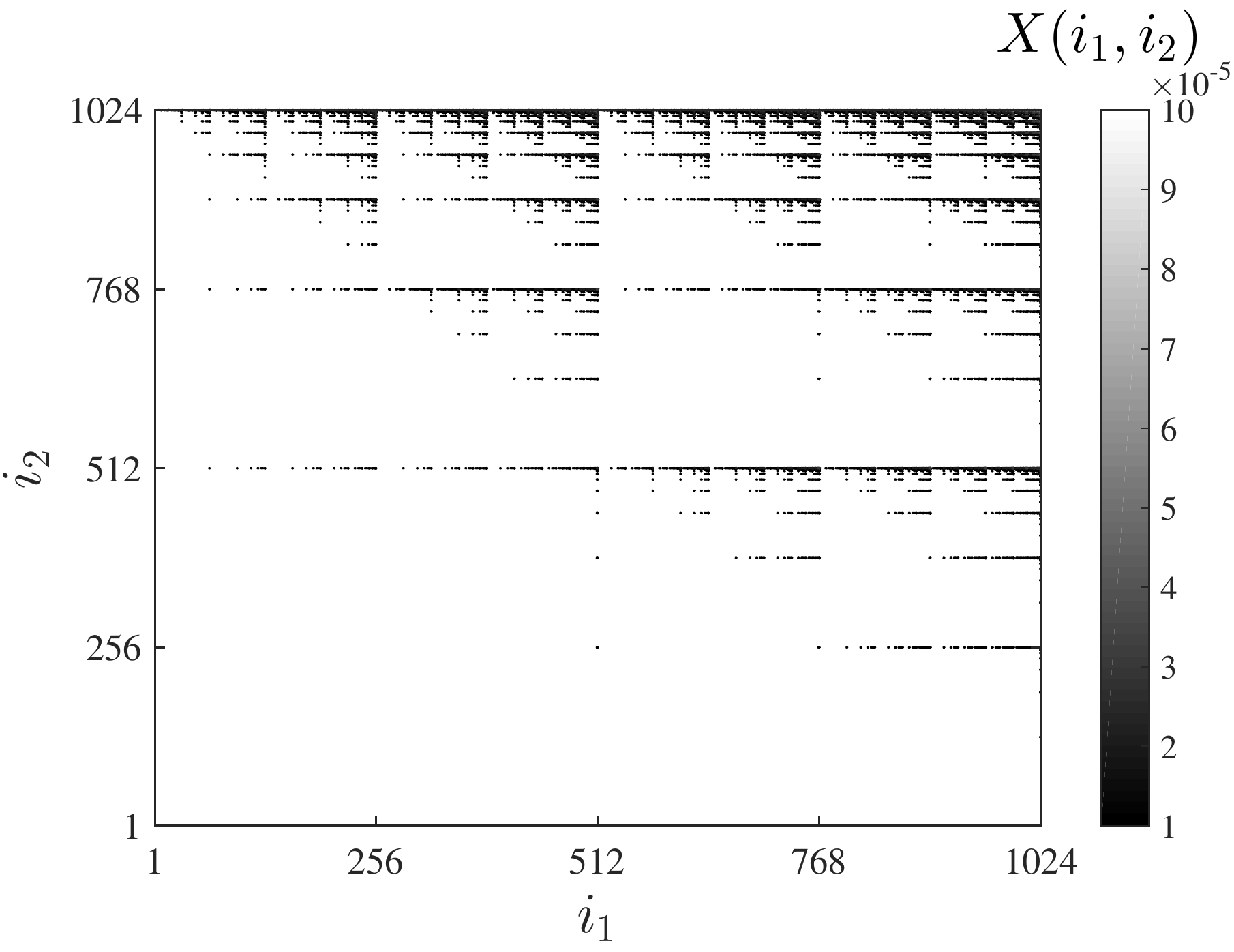}
  \caption{Two-dimensional multifractal measure with $p_1=0.1$, $p_2=0.2$, $p_3=0.3$, $p_4=0.4$ and size $1024\times1024$.}
  \label{Fig:TS:2D}
\end{figure}

In order to investigate the performance of the two-dimensional MF-DMA methods, we adopt the multiplicative cascading process to synthesize the two-dimensional multifractal measure. The process begins with a square, and we partition it into four sub-squares with the same size. We then assign four proportions of measure $p_1$, $p_2$, $p_3$ and $p_4$ to them (s.t. $p_1+p_2+p_3+p_4=1$). Each sub-square is further partitioned into four smaller squares and the measure is re-assigned with the same proportions. The procedure is repeated 10 times and we finally generate the two-dimensional multifractal measure with size $1024\times1024$. In Fig.~\ref{Fig:TS:2D}, the model parameters are $p_1=0.1$, $p_2=0.2$, $p_3=0.3$, and $p_4=0.4$. In this paper, we particularly adopt $\theta=\theta_1=\theta_2$ for the isotropic implementation of the two-dimensional MF-DMA analysis.

\begin{figure*}[htb]
  \centering
  \includegraphics[width=0.9\linewidth]{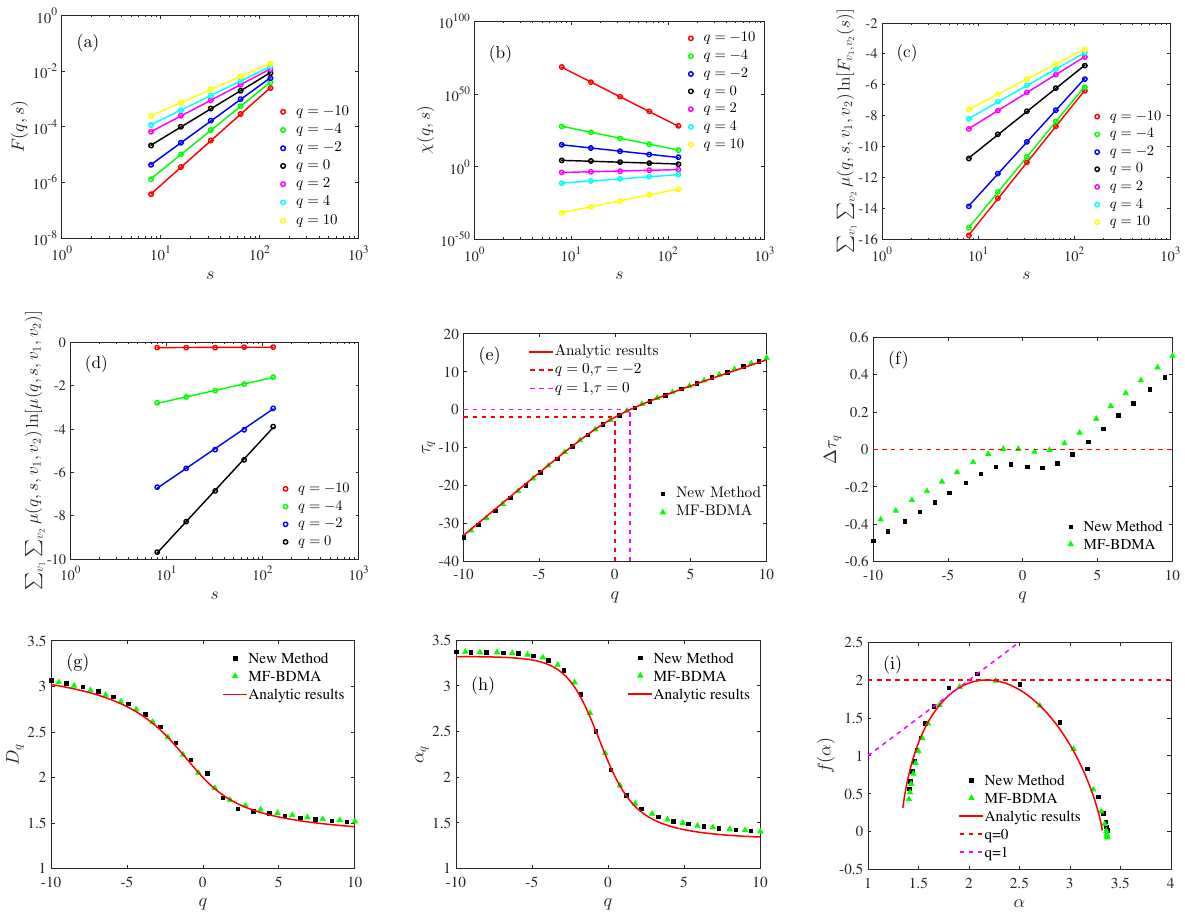}
  \caption{Multifractal analysis of the two-dimensional multifractal measure, based on the traditional MF-BDMA$(\theta_1,\theta_2,q)$ and the direct determination approach with $\theta_1=\theta_2=0$ (backward moving average). (a) Power-law dependence of $F(q,s)$ on box size $s$ for different $q$. (b) Power-law dependence of $\chi(q,s)$ on box size $s$ for different $q$. (c) Linear dependence of $\sum_{v_1}\sum_{v_2} \mu(q,s,v_1,v_2) \ln{F_{v_1,v_2}(s)}$ against $\ln{s}$. (d) Linear dependence of $\sum_{v_1}\sum_{v_2} \mu(q,s,v_1,v_2) \ln\left[\mu(q,s,v_1,v_2)\right]$ against $\ln{s}$. (e) The mass exponent function $\tau(q)$. (f) Differences $\Delta \tau(q)$ between the estimated mass exponents and their theoretical values. (g) The generalized dimensions $D_q$. (h) The singularity strength function $\alpha(q)$. (i) The multifractal singularity spectrum $f(\alpha)$.}
  \label{Fig:MF-DMA:2D:theta=0}
\end{figure*}

Fig.~\ref{Fig:MF-DMA:2D:theta=0} displays the backward case ($\theta_1=\theta_2=0$) of the two-dimensional MF-DMA analysis, using two approaches respectively. Fig.~\ref{Fig:MF-DMA:2D:theta=0}(a) illustrates the power-law dependence of the fluctuation function $F(q,s)$ on the scale $s$ for different $q$. The exponents $h(q)$ from the traditional MF-DMA method are obtained by the least squares fitting in log-log scales. Fig.~\ref{Fig:MF-DMA:2D:theta=0}(b) illustrates the power-law dependence of the partition function $\chi(q,s)$ on the scale $s$ for different $q$. The slopes obtained by linear regressions of $\ln \chi(q,s)$ against $\ln s$ are the estimates of $\tau(q)$, which are shown in Fig.~\ref{Fig:MF-DMA:2D:theta=0}(e). We obtain $D_q$ using $D_q = \frac{\tau(q)}{q-1}$ and $\alpha(q)$ and $f(\alpha)$ using the Legendre transform, which are presented in Fig.~\ref{Fig:MF-DMA:2D:theta=0}(g-i). On the other hands, the multifractal nature of the two-dimensional $p$-model can also be estimated by the direct determination approach.
Fig.~\ref{Fig:MF-DMA:2D:theta=0}(c) plots the dependence of $\sum_{v_1}\sum_{v_2} \mu(q,s,v_1,v_2) \ln{F_{v_1,v_2}(s)}$ against $s$ and Fig.~\ref{Fig:MF-DMA:2D:theta=0}(d) plots the dependence of $\sum_{v_1}\sum_{v_2} \mu(q,s,v_1,v_2) \ln\left[\mu(q,s,v_1,v_2)\right]$ against $s$ in linear-log coordinates. The slopes of the linear fits in Fig.~\ref{Fig:MF-DMA:2D:theta=0}(c) and Fig.~\ref{Fig:MF-DMA:2D:theta=0}(d) are the direct estimates of $\alpha(q)$ and $f(\alpha)$, which are shown in Fig.~\ref{Fig:MF-DMA:2D:theta=0}(h-i). In Fig.~\ref{Fig:MF-DMA:2D:theta=0}(e) and Fig.~\ref{Fig:MF-DMA:2D:theta=0}(g-i), we also show the analytical solution as a continuous curve for comparison. The analytical formula of $\tau(q)$ is expressed as following
\begin{equation}
  \tau_{\rm{analy}}(q)=-\frac{\ln(p_1^q+p_2^q+p_3^q+p_4^q)}{\ln2}.
  \label{Eq:MF-DMA:tau:analy:2D}
\end{equation}
We also show the analytical singularity spectrum as a continuous curve for comparison, where the singularity strength function $\alpha(q)$ can be calculated as follows
\begin{equation}
  \alpha_{\rm{analy}}(q)=-\frac{p_1^q\ln{p_1}+p_2^q\ln{p_2}+p_3^q\ln{p_3}+p_3^q\ln{p_3}}{(p_1^q+p_2^q+p_3^q+p_4^q)\ln2}.
  \label{Eq:alpha:analy:2D}
\end{equation}
The analytical expressions of $D_q$ and the multifractal spectrum $f(\alpha)$ can be obtained via $D_q = {\tau(q)}/{(q-1)}$ and $f(\alpha) = q\alpha - \tau(q)$ respectively. In Fig.~\ref{Fig:MF-DMA:2D:theta=0}(e), we mark $\tau(0)=-2$ (hence the fractal dimension $D_0=2$) and $\tau(1)=0$, while in Fig.~\ref{Fig:MF-DMA:2D:theta=0}(i), we show that $f^\prime(\alpha\mid q=0) = 0$ and $f^\prime(\alpha\mid q=1) = 1$.

We find the two-dimensional MF-DMA analysis has the same properties as the one-dimensional case in Fig.~\ref{Fig:MF-DMA:p-model:theta=0}. The estimation deviations $|\Delta\tau(q)|$ became bigger when the $q$ values are far from 0. Very negative $q$'s cause the mass exponents $\tau(q)$ underestimated and very positive $q$'s cause $\tau(q)$ overestimated. Correspondingly, when the $|q|$ values are large, the estimated $D_q$ in Fig.~\ref{Fig:MF-DMA:2D:theta=0}(g) and $\alpha(q)$ in Fig.~\ref{Fig:MF-DMA:2D:theta=0}(h) are slightly higher than their respective analytic values, while the estimated $f(\alpha)$ curves in Fig.~\ref{Fig:MF-DMA:2D:theta=0}(i) are slightly right-biased to their analytic values. Another property observed in Fig.~\ref{Fig:MF-DMA:2D:theta=0}(f) is that the $\Delta\tau(q)$ curves estimated by the direct determination approach locate lower than those by the traditional approach, which is more obvious than the one-dimensional case. This indicates that, with very positive $q$ values, the direct determination approach outperforms the traditional approach for the MF-DMA analysis. However, with very negative $q$ values, the traditional approach performs better.

\begin{figure*}[tb]
  \centering
  \includegraphics[width=0.9\linewidth]{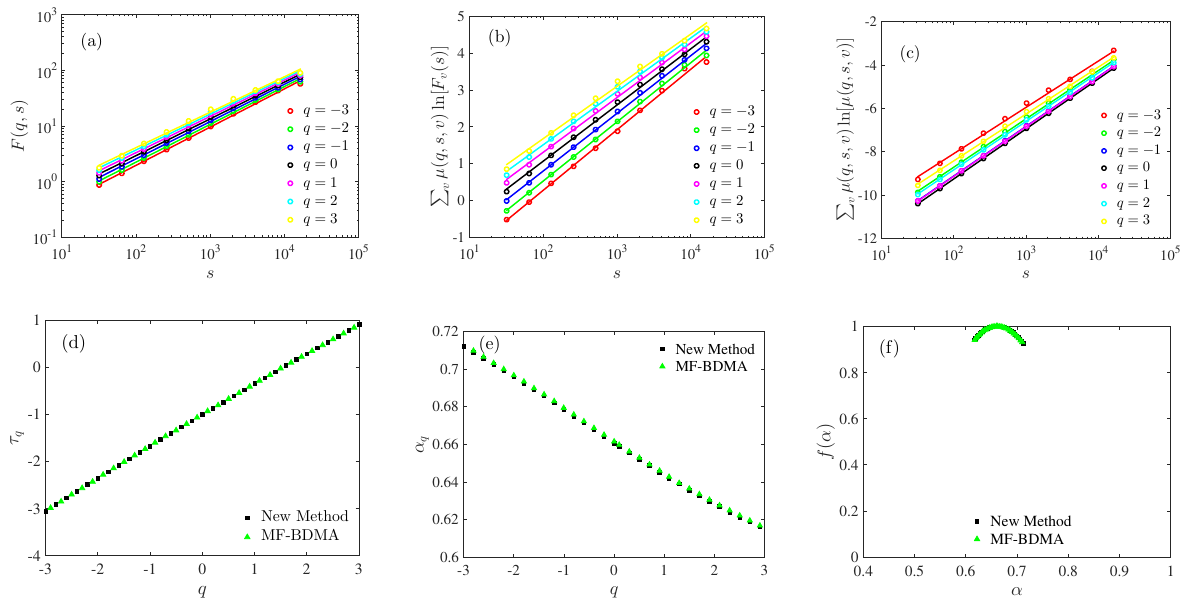}
  \caption{Multifractal analysis of fractional Brownian motions with $H_{in}=0.7$, based on the traditional MF-BDMA$(\theta,q)$ and the direct determination approach with $\theta=0$. (a) Power-law dependence of $F(q,s)$ on box size $s$ for different $q$. (b) Linear dependence of $\sum_v \mu(q,s,v) \ln{F_v(s)}$ against $\ln{s}$. (c) Linear dependence of $\sum_v\mu(q,s,v)\ln[\mu(q,s,v)]$ against $\ln{s}$. (d) The mass exponent function $\tau(q)$. (e) The singularity strength function $\alpha(q)$. (f) The multifractal singularity spectrum $f(\alpha)$. }
  \label{Fig:MF-DMA:fbm:theta=0}
\end{figure*}

\subsection{Fractional Brownian motion}

\begin{figure*}[htb]
  \centering
  \includegraphics[width=0.9\linewidth]{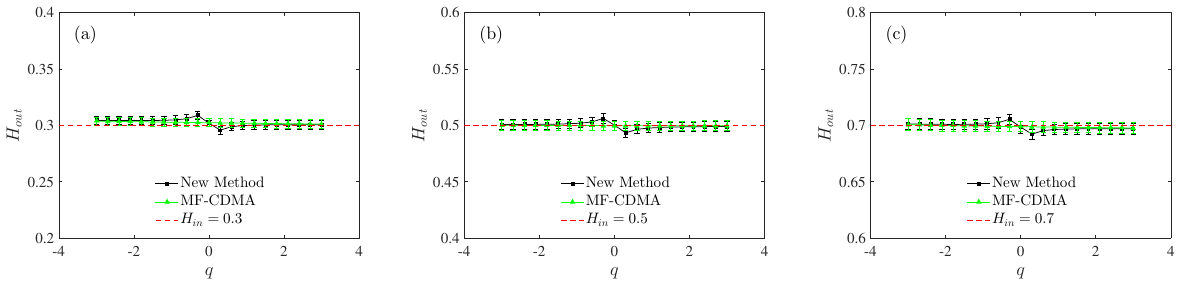}
  \caption{Hurst exponent estimates. In each plot, two types of markers are obtained from the direct determination approach and the original approach with $\theta=0.5$. Each point shows the average Hurst index estimated over 100 simulated time series. The error bars show the standard deviations. Each column corresponds to a fixed Hurst index ($H_{in} = 0.3, 0.5$ and 0.7 from left to right).}
  \label{Fig:MF-DMA:fbm:Hurst}
\end{figure*}

We also test the performance of the new method using monofractal time series. Fig.~\ref{Fig:MF-DMA:fbm:theta=0} shows the results of multifractal analysis on fractional Brownian motions (FBM), using the direct determination approach and the traditional approach. FBM series are generated by using a wavelet-based generator (WFBM) \cite{Abry-Sellan-1996-ACHA} with input Hurst exponent $H_{in}=0.7$. We take the backward case (MF-BDMA) to present the results. Fig.~\ref{Fig:MF-DMA:fbm:theta=0}(a) illustrates the power-law dependence of the fluctuation function $F(q,s)$ on the scale $s$ for different $q$'s. We notice that these lines almost have the same slopes. In other words, the estimated exponents $h(q)$ according to Eq.~(\ref{Eq:MF-DMA-OG:Fqs:h}) are all close to $H_{in}=0.7$. This results in an almost linear $\tau(q)$ function in Fig.~\ref{Fig:MF-DMA:fbm:theta=0}(d) and an almost linear $\alpha(q)$ function in Fig.~\ref{Fig:MF-DMA:fbm:theta=0}(e), obtained respectively from Eq.~(\ref{Eq:MF-DMA-OG:tau:hq}) and Eq.~(\ref{Eq:MF-DMA-OG:alpha}). The strength of the multifractal nature can be quantified by the width of the singularity spectrum $\Delta\alpha = \alpha_{\max}-\alpha_{\min}$. In Fig.~\ref{Fig:MF-DMA:fbm:theta=0}(f), we illustrate the function $f(\alpha)$ as a function of $\alpha$ and find that the spectrum width is very narrow. This confirms that the fractional Brownian motion signal is monofractal, not multifractal.

On the other hands, the spurious multifractal nature for fractional Brownian motion can also be estimated by the direct determination approach. In Fig.~\ref{Fig:MF-DMA:fbm:theta=0}(b,c) we present $\sum_v \mu(q,s,v) \ln{F_v(s)}$ and $\sum_v\mu(q,s,v)\ln[\mu(q,s,v)]$ as a function of the time lag $s$. We find these lines almost have the same slopes. Hence, both the singularity strength $\alpha(q)$ in Fig.~\ref{Fig:MF-DMA:fbm:theta=0}(e) and the multifractal singularity spectrum $f(\alpha)$ in Fig.~\ref{Fig:MF-DMA:fbm:theta=0}(f) have narrow domains. In Fig.~\ref{Fig:MF-DMA:fbm:theta=0}(d-f), we show that the direct determination approach performs comparable to the traditional approach.

\begin{figure*}[htb]
  \centering
  \includegraphics[width=0.9\linewidth]{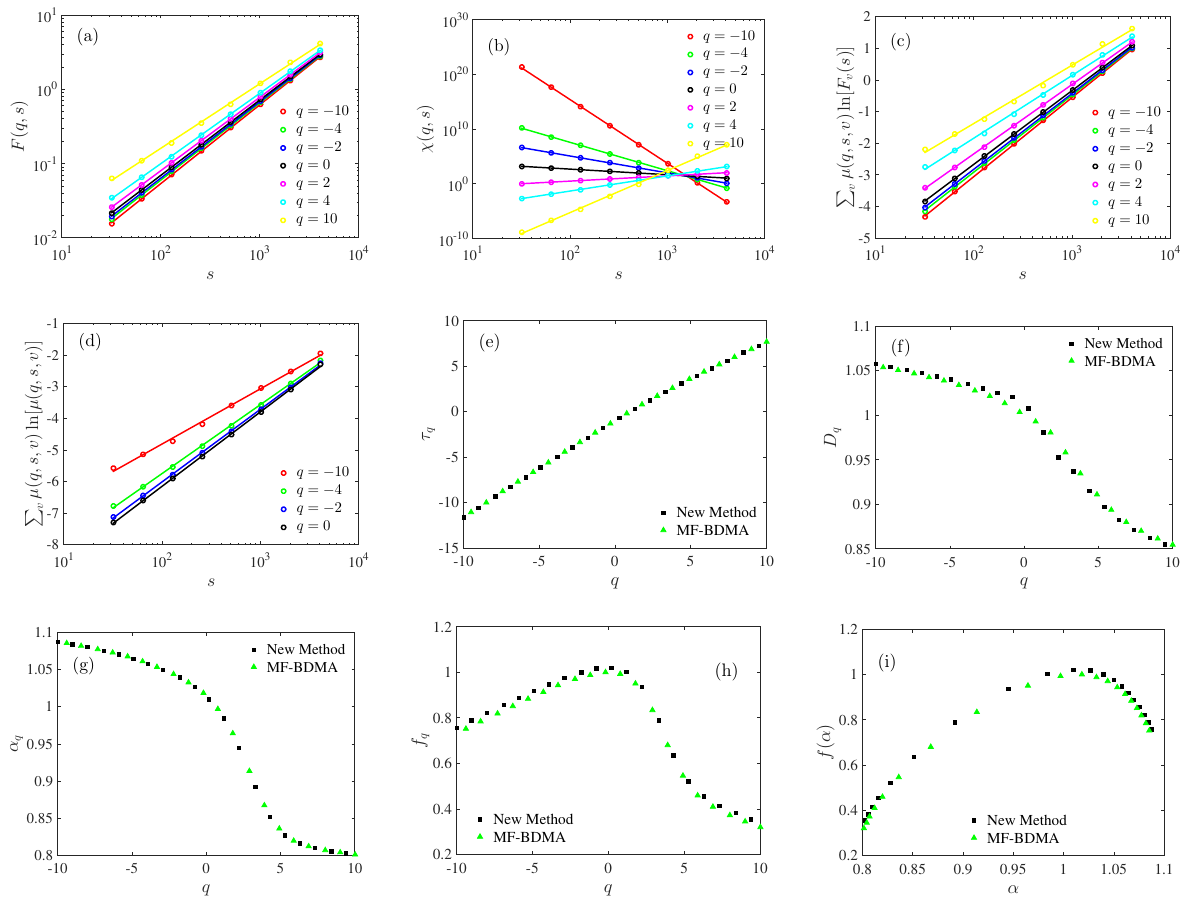}
  \caption{Multifractal analysis of the 1-minute volatility time series of SPD Bank (600000) stock price, based on the traditional MF-BDMA$(\theta,q)$ and the direct determination approach with $\theta=0$ (backward moving average). (a) Power-law dependence of $F(q,s)$ on box size $s$ for different $q$. (b) Power-law dependence of $\chi(q,s)$ on box size $s$ for different $q$. (c) Linear dependence of $\sum_v \mu(q,s,v) \ln{[F_v(s)]}$ against $\ln{s}$. (d) Linear dependence of $\sum_v\mu(q,s,v)\ln[\mu(q,s,v)]$ against $\ln{s}$. (e) The mass exponent function $\tau(q)$. (f) The generalized dimensions $D_q$. (g) The singularity strength function $\alpha(q)$. (h) The $f(q)$ function. (i) The multifractal singularity spectrum $f(\alpha)$.}
  \label{Fig:MF-DMA:Return:theta=0}
\end{figure*}

We also compare the estimated accuracy of Hurst exponent by two approaches in Fig.~\ref{Fig:MF-DMA:fbm:Hurst}. With the traditional approach, the estimated Hurst exponent can be obtained by Eq.~(\ref{Eq:MF-DMA-OG:Fqs:h}). With the new method, we first get $\tau(q)$, then backward derive $h(q)$ through Eq.~(\ref{Eq:MF-DMA-OG:tau:hq}). Note that $h(0)=\lim_{q\to0}\frac{\tau(q)+1}{q}=\tau^{\prime}(0)$. We generate FBM time series with three different input Hurst indexes ($H_{\rm{in}} = 0.3$, 0.5 and 0.7). For each $H_{\rm{in}}$, we simulate 100 FBM time series. We present the results of centered detrending moving average case with $\theta=0.5$ (CDMA), since CDMA has the best performance \cite{Shao-Gu-Jiang-Zhou-Sornette-2012-SR}. We confirm that both the direct determination approach and the traditional approach can produce relatively accurate estimation for all the three different $H_{in}$ cases. In addition, the estimates show no obvious difference except when $q$ is close to 0. With canonical approach, slight numerical errors or approximations in $\tau(q)$ will cause big fluctuations in $h(q)$ when $q\to 0$.

\section{Application to financial time series}
\label{S1:Application}

We now apply the direct determination approach and the traditional approach to investigate the multifractal properties of the volatility time series of SPD Bank (600000) stock price. The volatility is defined as the absolute value of the logarithmic difference of 1-min closing prices:
\begin{equation}
  R(t)=|\ln P(t)-\ln P(t-1)|,
  \label{Eq:MF-DMA:return}
\end{equation}
where $P(t)$ is the closing price on minute $t$. The time period of the samples is from 5 January 2015 to 14 March 2016, containing 70,179 data points.

Here we take the backward method for example. Fig.~\ref{Fig:MF-DMA:Return:theta=0}(a) shows the power-law dependence of the fluctuation function $F(q,s)$ on the scale $s$. For different $q$'s, the slops $h(q)$ are different, but the disparity is not as obvious as that of the $p$-model. For the new method, the partition function $\chi(q,s)$ scales with respect to the scale $s$ as a sound power law in Fig.~\ref{Fig:MF-DMA:Return:theta=0}(b). The slops are the estimated $\tau(q)$ which are almost overlapping with the estimation by traditional method, as shown in Fig.~\ref{Fig:MF-DMA:Return:theta=0}(e). The singularity strength function and the multifractal spectrum are also directly determined. Fig.~\ref{Fig:MF-DMA:Return:theta=0}(c) shows a sound linear dependence of $\sum_{v=1}^{N_s} \mu(q,s,v) \ln F_v(s)$ against $\ln s$ and Fig.~\ref{Fig:MF-DMA:Return:theta=0}(d) shows a sound linear dependence of $\sum_{v=1}^{N_s} \mu(q,s,v) \ln\left[\mu(q,s,v)\right]$ against $\ln s$. The slops of these lines are different such that the $\alpha(q)$ values range from 0.80 to 1.09 and $f(\alpha)$ range from 0.34 to 1 in Fig.~\ref{Fig:MF-DMA:Return:theta=0}(g-i). The strength of multifractality can be characterized by the span of the multifractal singularity strength function. Therefore, We observe in the figure that the 1-min volatility time series of SPD Bank possesses multifractal nature, and that both the direct determination approach and the traditional approach show identical results.

\section{Conclusions}
\label{S1:Conclude}

In this paper, we defined a canonical measure to develop a new detrending moving average approach for multifractals such that the singularity strength function $\alpha(q)$ and the multifractal spectrum $f(\alpha)$ can be directly determined. In the canonical framework, the mass scaling exponent $\tau(q)$ is defined via the partition function. This is a different statistical approach compared with the traditional MF-DMA approach, which computes the generalized Hurst exponent directly and deduces the singularity spectrum indirectly.

We focused on the multifractal analysis in one and two dimensions. Extensions to higher dimensions are straightforward. The performances of the direct determination approach and the traditional approach are tested based on synthetic multifractal measures with known theoretical multifractal properties, including one-dimensional $p$-model, two-dimensional $p$-model and fractional Brownian motion.

We found that the direct determination approach has comparable performance with the traditional approach to unveil the multifractal nature. In other words, for the $p$-model, both approaches provide results agreeing with the analytical expressions of the mass function, the singularity strength function and the multifractal spectrum. In more careful comparisons, for the backward MF-DMA case with $\theta=0$ and the forward MF-DMA case with $\theta=1$, when $q$ is very positive, the direct determination approach performs slightly better; when $q$ is very negative, the traditional approach performs slightly better. For the centered MF-DMA case with $\theta=0.5$, these two approaches do not show obvious differences. Both two approaches confirm that fractional Brownian motion signals are monofractal. For the estimates of Hurst exponent, the canonical approach performs slightly worse than the traditional approach when $q$ is close to 0, because the Hurst exponent is deduced indirectly with the traditional approach. Finally, when the new approach is applied to the 1-min volatility time series of stock prices, the multifractal nature is confirmed.

In all, we contribute to extend the direct determination approach of singularity spectrum \cite{Chhabra-Jensen-1989-PRL} to the detrending moving average analysis. Of course, this direct determination approach is also possible for other detrending analysis, such as multifractal detrended fluctuation analysis (MF-DFA). 


\begin{acknowledgments}

We acknowledge financial support from the National Natural Science Foundation of China (71501072, 71671066, 71532009, 11375064) and the Fundamental Research Funds for the Central Universities (222201524004, 222201718006).

\end{acknowledgments}


%


\end{document}